\def\E {{\em Einstein}}
\def\cq {$\chi^2$}
\def\simlt {\lower.5ex\hbox{$\; \buildrel < \over \sim \;$}}
\def\simgt{\lower.5ex\hbox{$\; \buildrel > \over \sim \;$}}

\def\net {$n_{\rm e}t$}

\def\sqrcm  {cm$^{-2}$}

\documentstyle[aabib, epsfig]{l-aa-ps}
\begin{document}
\thesaurus{09.01.1; 09.09.1 N132D; 09.19.2; 13.25.4}

\title{BeppoSAX LECS/MECS X-ray spectroscopy of the young supernova remnant
  N132D}

\author{F. Favata\inst{1}  \and J. Vink\inst{2} \and A.\,N. Parmar\inst{1}
  J.\,S. Kaastra\inst{2} \and T. Mineo\inst{3}}

\institute{Astrophysics Division -- ESA/ESTEC,
 Postbus 299, NL-2200 AG Noordwijk, The Netherlands
\and
Space Research Organization Netherlands, Sorbonnelaan 2, NL-3584 CA,  
Utrecht, The Netherlands 
\and
Istituto di Fisica Cosmica ed Applicazioni Informatiche, CNR, Via
 U. La Malfa 153, I-90146 Palermo, Italy 
}

\offprints{F. Favata (fabio.favata@astro.estec.esa.nl)}

\date{Received date ; accepted date}

\maketitle 
\begin{abstract}

  We have analyzed an observation of the supernova remnant N132D in
  the Large Magellanic Cloud, performed with the Low Energy
  Concentrator Spectrometer (LECS) and the Medium Energy Concentrator
  Spectrometer (MECS) instruments on board the X-ray satellite
  BeppoSAX. Thanks to the good sensitivity of the LECS/MECS
  combination over the range 0.3--8.5~keV we were able to detect the
  presence of an additional hotter plasma with respect to what had
  been previously assumed for this remnant. This hotter component is
  the source of the Fe-K line emission visible in the spectrum. Using
  a two component NEI plasma model we find that the best fitting
  abundance values indicate that N132D is also O-rich from the X-ray
  point of view.  We briefly discuss the origin of the hot component
  and tentatively associate it with a radio bright region.

  \keywords{ISM: abundances; ISM: individual objects: N132D: ISM:
    supernova remnants: X-rays: ISM }

\end{abstract}

\section{Introduction}
\label{sec:intro}

N132D is a radio and X-ray bright, relatively young ($\simeq
3000$\,yr, \cite{mwk95}) supernova remnant (SNR) located in the Large
Magellanic Cloud (LMC). The optical spectrum of the ejecta shows
evidence for a very high oxygen abundance, as well as for large
expansion velocities. This indicates that N132D belongs, like the well
known galactic remnant Cas~A, to the class of O-rich remnants. Even at
the distance of the LMC ($\simeq 50$\,kpc), the soft X-ray and UV flux
at Earth of N132D is relatively large, and as a consequence the
remnant has been observed by many low-energy X-ray instruments, both
with dispersive (the \E\ Focal Plane Crystal Spectrometer --- FPCS)
and non-dispersive spectrographs (e.g. the \E\ Solid State
Spectrometer --- SSS, the ROSAT PSPC and the ASCA SIS). A
comprehensive analysis of all the X-ray spectra from the different \E\ 
instruments has been presented by \cite*{hhc+93} --- hereafter H93
---, who have made full use of the relatively high spectral resolution
of the SSS. They have shown that the spectrum, at the SSS resolution,
is well modeled with a relatively cool single-temperature
non-equilibrium ionization (NEI) model, and have determined best-fit
abundance values and ionization time scales for the remnant.

While the \cq\ of the single-component H93 model is satisfactory,
their result has one remarkable feature: the derived best-fit metal
abundances for the plasma are lower than the mean abundances for the
LMC. While detailed abundance studies for the local neighborhood of
N132D have not been performed, such a low abundance is surprising,
since the emission would be expected to come from a mixture of the
supernova ejecta, heated by the reverse shock and of the shocked
interstellar or circumstellar medium (ISM/CSM), similar to what has
been observed for Cas~A itself. The fast moving O-rich knots observed
in the optical and UV (\cite{brl94}) are obviously pure ejecta
components, showing the products of recent nucleo-synthesis. It is
thus unclear how a mixture of plasmas with high (the ejecta) and local
(CSM/ISM) abundances can produce an X-ray spectrum with abundances
lower than the local value (ISM).
 
Given the fact that N132D is one of the brightest X-ray sources in the
LMC and that it has a small spatial extent (with an X-ray diameter of
$\simeq 110$\,arcsec in the \E\ HRI images), it was chosen as a
calibration target for the Science Verification Phase (SVP) of the
BeppoSAX program (\cite{bbp+97}). In this paper we discuss the X-ray
spectrum of N132D obtained with the Low Energy Concentrator
Spectrometer (LECS; \cite{pmb+97}) and the Medium Energy Concentrator
Spectrometers (MECS; \cite{bcc+97}) on board BeppoSAX. The good
spectral resolution of the LECS (comparable to the resolution of CCD
detectors at energies below 0.5\,keV) and its good sensitivity down to
0.1~keV, in combination with the high sensitivity of the MECS above
2~keV the broad band X-ray spectrum of N132D to be studied, from the
low-energy interstellar cut-off at $\simeq 0.3$\,keV up to $\simeq
8$~keV.

The combined LECS/MECS spectrum of N132D clearly shows the presence of
the Fe\,K complex at $\simeq 6.7$\,keV, implying that some of the
plasma has a temperature higher than found from the analysis of the
\E\ data, making it necessary to fit the spectrum with two plasma
components. As will be shown, this results in higher metal abundances
for the best-fit model, which remove the apparent discrepancy with LMC
ambient abundances.

\section{Data reduction}

The BeppoSAX SVP observation of N132D took place on November 22, 1996,
and it resulted in 15\,ks of effective exposure time in the LECS and
33\,ks in the MECS (the difference being due to the LECS being
operated during Earth night only). The LECS data were reduced through
the LECS pipeline software (SAX-LEDAS 1.4.0), while the MECS data were
reduced using the XAS V.\ 2.0.1 package. The reduction was performed
separately for each unit. The source spectra were extracted from a
circle centered on the source itself of radius 8\,arcmin and 4\,arcmin
respectively for the LECS and the MECS. Publicly available detector
response matrices (known as ``December 31, 1996'' issue) were used for
the MECS, while the LECS response matrix was computed with the LEMAT
package (V.\ 3.2.0).  To subtract the background the standard LECS and
MECS background files obtained by adding up a set of ``empty sky''
observations were used, extracted from the same circular regions as
the source spectrum.  The source spectrum was re-binned so to have at
least 20 counts per re-binned channel. For the LECS, channels with
energies between 0.3 and 5.0\,keV were retained for the spectral
analysis, while the range was 1.6--8\,keV for the MECS. The resulting
(background-subtracted) source count rate is 0.96\,cts\,s$^{-1}$ for
the LECS and 0.42\,cts\,s$^{-1}$ for the MECS. Given the current
uncertainties on the absolute calibration of the various detectors,
the relative normalization of the MECS detectors with respect to the
LECS has been left as an additional free parameter in the fit. The fit
converges to a relative normalization of the MECS detectors $\simeq
25$\% higher than the LECS, in line with the expected calibration
uncertainties at this stage and with the results from other SVP
targets.

\section{Spectral analysis}

The spectral analysis was performed using the SRON SPEX (V.\ 1.10)
package, which contains state-of-the-art plasma emission models as
well as NEI models which are well suited to the analysis of the X-ray
emission from supernova remnants (\cite{kmn95}).

\begin{figure}[htbp]
  \begin{center}
    \leavevmode
    \epsfig{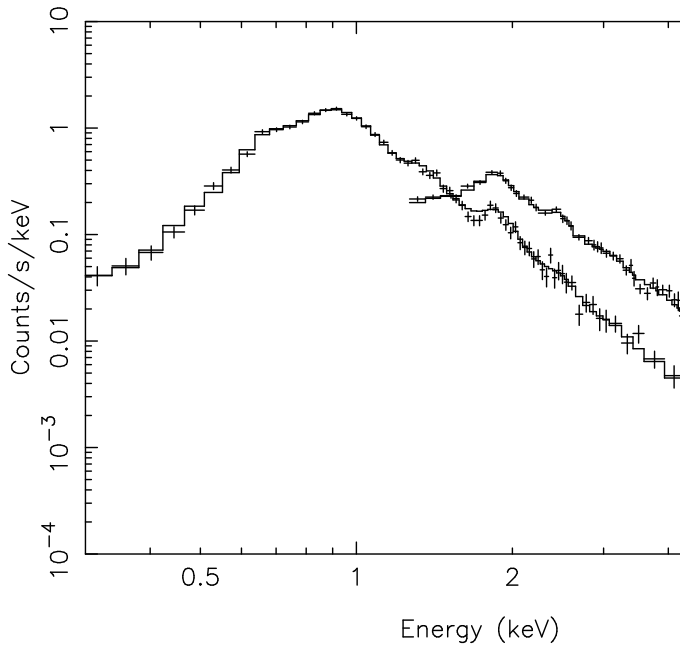} \end{center}
  \caption{The observed BeppoSAX LECS and MECS spectra of N132D,
  together with the best-fit two-component NEI model described in the
  text.}  \label{fig:spec}
\end{figure}

The LECS spectrum alone, given its low signal-to-noise at the higher
energies, allows a reasonable fit to the data with an isothermal NEI
model with a temperature around 1\,keV and metal abundances compatible
with H93, although the harder part of the spectrum is not very well
described by this fit and the LECS data alone already suggest the
presence of a Fe\,K line. This is confirmed by the MECS spectrum, in
which a line at $\simeq 6.7$\,keV is evident, implying emission due to
the Fe\,K complex (see Fig.~\ref{fig:spec}), and indicating the
presence of plasma at higher temperatures. Detailed comparison of the
ROSAT HRI X-ray images of the remnant with narrow-band HST optical
images (\cite{mbd+96}) show that no X-ray emission is coming from the
expanding ejecta inside the shell (which are well visible in the HST
images), and that the X-ray emission is rather associated with the
swept-up circumstellar medium (which may however be mixed with some
ejecta material). In this regard N132D is thus different from the
Cas~A remnant.

To fit the combined LECS/MECS spectra we have therefore used a
two-temperature NEI model, in which the abundance of O, Ne, Mg, Si, S,
Ar and Fe are allowed to individually vary. As we assume that both
components originate in the swept-up CSM, the abundances for each
element were coupled across the two components. The other free
parameters were the interstellar column density and the emission
measure, temperature and ionization parameter (\net) for each
component. The last parameter indicates the ionization stage of a
plasma of density $n_{\rm e}$ at time $t$ after it was shocked to the
plasma temperature.  Note that in reality temperature and density may
have varied since the first time the plasma was shocked. The
abundances of He, C, N, Ca and Ni were set to the LMC values
(\cite{rd92}). None of the elements whose ratios have been constrained
a priori have sufficiently strong or well resolved lines in the X-ray
spectrum to allow their determination to be derived from the spectrum
itself. At the same time, their abundance cannot be allowed to freely
vary, as they would influence the rest of the abundances in
non-physical ways. In particular Ca and Ni are in such a relatively
soft spectrum more constrained by there L-shell lines below 1~keV than
by their K-shell emission at higher energy (this is always true for
Ni). There, they tend to fill up gaps in the model, arising from
uncertainties in instrumental response and in the plasma emission code
(e.g. in the Fe L-shell emission) or from statistical noise. For the
interstellar absorption we used the \cite*{mm83} model.

The best-fit parameters for the two temperature NEI model are shown in
Table~\ref{tab:fitpar}. This model, which yields a reduced \cq\ of
1.1, is plotted, together with the source spectrum, in
Fig.~\ref{fig:spec}. Note that the best-fit parameters do not
significantly depend on the assumed C and N abundances, which are
fixed here to the LMC canonical values ($0.52$ and $0.37$ times solar,
respectively).  Assuming values in the range 0.1--1.0 times the solar
value does not significantly influence the fit.  The best-fit
abundances thus obtained are in line with the canonical LMC abundances
derived from optical spectroscopy (with the exception of Si whose
X-ray determined abundance is still slightly lower than the canonical
LMC value), showing that the SNR is not metal-deficient with respect
to the ambient medium.

The higher best-fit metal abundances with respect to the
single-component NEI model are linked to the addition of the hotter
plasma component. Among other effects, this induces a lower predicted
line-to-continuum ratio around the O\,{\sc viii} line complex, because
for the hot plasma component O is almost completely ionized. The same
is more or less true for Ne and Mg.  Another important difference with
respect to the single-component model is that we find a best fit
(though not very well constrained) interstellar column density $N_{\rm
  H} = 3\times 10^{21}$~\sqrcm, whereas H93 obtained a best fit value
$N_{\rm H} = 0.6\times 10^{21}$~\sqrcm.  A higher column density
implies more emission below 1\,keV, and since the continuum is, in
such an absorbed X-ray spectrum, mainly determined by line-free
regions above $\simeq 3$\,keV, the extra soft emission needed to
compensate the higher absorption has to come from line emission of
mainly O\,{\sc viii} and Fe-L.

At a distance of the LMC of $\simeq 50$\,kpc the diameter of N132D
($\simeq 110$\,arcsec) corresponds to a radius of 13.5\,pc, implying
an emitting volume $V = f\cdot 2.9 \times 10^{59}$\,cm$^3$, where $f$
is the filling factor. A total emission measure at 50\,kpc of
$175\times 10^{58}$\,cm$^{-3}$ implies then a mass $M = f^{0.5}\cdot
760\,M_\odot$. The canonical filling factor for a strong shock is
0.25, although the ``clumping'' of the soft X-ray emission seen in the
ROSAT HRI images (\cite{mbd+96}) would indicate a smaller filling
factor. On the other hand, \cite*{dm95b} claim, based on the shell
thickness at radio wavelength, that a filling factor of 0.5 may be a
more appropriate value, although the relevance of the radio-determined
filling factor to the X-ray emission is not obvious. If we thus assume
a possible range of 0.1 to 0.5 for the filling factor (with an extra
15\% uncertainty for the size of the remnant) the derived total mass
for the N132D X-ray emitting material is between 250 and
600\,$M_\odot$, a large value which confirms that the X-ray emission
is mostly associated with swept-up CSM/ISM and not with stellar
ejecta. The pre-shock density inferred is close to 1\,cm$^{-3}$. The
$\simeq 450\,M_\odot$ of shocked CSM/ISM contain, assuming that the
matter has the canonical LMC abundances of \cite*{rd92}, about
$6\,M_\odot$ of O; at the same time the high O abundance of the X-ray
spectrum (significantly higher than the LMC ambient O abundance)
implies a total O content for the X-ray emitting plasma a few
$M_\odot$ higher, consistent with the X-ray emitting material
containing a small amount of mixed ejecta, and with N132D being an
O-rich remnant, having its origin in a core collapse supernova.

\begin{table}[htbp]
\caption{The best-fit parameters for the two-temperature NEI model
  fit to the LECS/MECS X-ray spectrum of N132D. Also shown, for each
  parameter, the formal 90\% range (i.e. $\Delta \chi = 2.71$). All
  the parameters are derived assuming a distance to the SNR of
  50\,kpc.  All abundances are relative to the solar value. The mean
  values for the LMC abundances are from Russell \& Dopita (1992).}
\label{tab:fitpar}
\begin{flushleft}
\begin{tabular}{llll} \hline \\[-10pt]
Parameter & Best-fit& 90\% range & LMC\\
\hline \\[0.1pt]
$N_H$ ($10^{21}$\,cm$^{-2}$)    & 3.0 & 0.9 -- 6.5 & \\[1pt]
{\bf component 1} & & &\\
$kT$ (keV)             & 0.8  & 0.55--0.85 & \\
$n_e\,n_H\, V$ ($10^{58}$\,cm$^{-3}$) & 158 & 107 -- 240  & \\
$n_e\,t$ ($10^3$\,cm$^{-3}$\,yr) & 2.3  & 0.5 -- 4.8 & \\[1pt]
{\bf component 2} & & &\\ 
$kT$ (keV)             & 2.7  & 1.8 -- 6.0 & \\
$n_e\,n_H\, V$ ($10^{58}$\,cm$^{-3}$) & 17 & 10 -- 32 & \\
$n_e\,t$ ($10^3$\,cm$^{-3}$\,yr) & 5.3  & 1.9 -- $\infty$& \\[1pt]
\noalign{\smallskip}
\mbox{[O/H]}  & 1.2 & 0.8 -- 5.0 & 0.32\\
\mbox{[Ne/H]} & 1.1 & 0.5 -- 3.5 & 0.42\\
\mbox{[Mg/H]} & 0.85 & 0.8 -- 1.8 & 0.74\\
\mbox{[Si/H]} & 0.62 & 0.4 -- 1.1 & 1.7\\
\mbox{[S/H]}  & 0.75 & 0.5 -- 2.0 & 0.27\\
\mbox{[Ar/H]} & 0.73 & 0.0 -- 2.5 & 0.49\\
\mbox{[Fe/H]} & 0.52 & 0.35 -- 0.75 & 0.50\\
\hline \\[0.1pt]
\end{tabular}
\end{flushleft}
\end{table}

The discovery of a hot plasma component in the X-ray spectrum of N132D
raises the question of the origin of this temperature structure. The
usual interpretation of a two-component NEI model, i.e.\ that the
low-temperature component corresponds to the reverse shock going
through the ejecta and that the hot component corresponds to the
shock-heated swept-up circumstellar mass, does not apply here, as
shown both by the HRI/HST comparison discussed above and by the high
implied X-ray emitting mass, higher than any plausible stellar mass.
Even associating the hot component only with pure ejecta would still
result in too high a mass, given also the predominance of non-X-ray
emitting ejecta in the remnant. It is thus not possible to separately
derive a mass for the ejecta from the X-ray spectrum.

The high mass of this remnant suggests that it is in, or entering the
Sedov phase of its evolution. Models for the temperature structure of
a pure Sedov shock model show that there is a gradient in the
temperature, i.e.\ a hot tenuous medium is present inside the swept-up
shell. The Sedov model (as implemented in SPEX) produces an acceptable
fit to the data, but with a lower best-fit interstellar column density
and a lower O abundance than implied by the two-component NEI model.
The derived size and age of the remnant and the density of the ISM
(\cite{kj93}) appear however to be an order of magnitude larger than
the actual values, casting doubts on the actual applicability of a
pure Sedov model.

A more promising explanation for the presence of the hot component
(which has significantly lower emission measure than the cool
component) involves the inhomogeneity of the CSM/ISM. A shock
propagating through an inhomogenuous medium will have different shock
velocities in different regions. Where the density is higher the shock
decelerates more rapidly, resulting in a cooler post-shock
temperature, proportional to the square of the velocity.  The
temperature of the cool component corresponds to a shock velocity
close to 600\,km\,s$^{-1}$, whereas the temperature of the hot
component indicates velocities around 1100\,km\,s$^{-1}$. The latter
value is still slower than the velocity of the presumably freely
expanding O-rich knots (\cite{mbd+96}). The above velocities are
computed assuming that ion and electron temperatures are in
equilibrium; if this assumption is not valid, the actual velocities
can be higher.

If the hot component is due to the inhomogeneity of the CSM/ISM, can
this be localized? Given the relatively small size of the N132D
remnant when compared with the spatial resolution of the LECS and MECS
instruments, and the variation of the point spread function of the
detectors with energy, an investigation of the presence of possible
spectral variations with position requires a detailed analysis, which
will be the subject of a future paper. We have however performed a
preliminary analysis of the ROSAT PSPC data for N132D (the higher
on-axis spatial resolution and smaller spectral band of the PSPC make
such an analysis more straightforward), obtained from the public
archive, extracting two separate images, one in the 0.5--1.0\,keV band
and the other in the 1.3--2.1\,keV band, and calculating a
hardness-ratio image. The hardness-ratio map shows the presence of a
hard feature in the south-east region of the remnant, which could be
the seat of the hot plasma seen in the LECS/MECS spectra. This X-ray
hard region has approximately the same location as a radio-bright
region in the 3.5\,cm map of \cite*{dm95b}, again suggestive of its
possible association with the hotter plasma component.

\section{Conclusions}

The good spectral resolution of the LECS and MECS detectors over a
broad spectral range (here 0.3--8.5\,keV) allows a good determination
of the temperature structure and the interstellar column density for
the X-ray spectra of N132D, both of which are important for
determining abundances. The most relevant result obtained from the
analysis of the SVP X-ray BeppoSAX spectra of N132D is the detection
of a plasma component at hotter temperature than previously reported
for this remnant. The presence of the hotter plasma, which is evident
in the spectrum through its well-resolved Fe\,K line emission, makes
it necessary to use two NEI components to properly fit the data; given
the association of both components with the swept-up CSM (with some
ejecta mixed in) the abundances in the two components have been
coupled together. The best-fit parameters for the two-component model
imply abundances similar to (or in some cases higher than) the
canonical LMC abundances, thus showing that the puzzling
under-abundances reported by H93 on the basis of the \E\ SSS data are
likely to be the result of the neglect of the hotter component. Our
analysis confirms the high X-ray emitting mass of N132D, indicating a
SNR in or entering the Sedov phase of its evolution. In a future paper
we will perform a detailed comparative spatial analysis of the LECS,
MECS and ROSAT PCPS data.

\acknowledgements{The BeppoSAX satellite is a joint Italian and Dutch
  program. We thank S. Molendi for the useful discussions. We have
  made use of the HEASARC X-ray archive. This work was financially
  supported by NWO, the Netherlands Organization for Scientific
  Research. }

\end{document}